\documentclass[12pt]{iopart}
\usepackage{epsfig}
\begin{document}

\title[ ]{Neutrino Factories: Physics Potential}

\author{Steve Geer\dag\
\footnote[3]{sgeer@fnal.gov. Presented at NUFACT02, London, 1-6 July 2002}}

\address{\dag\ Fermi National Accelerator Laboratory, P.O. Box 500, 
Batavia, Illinois 60510, USA}

\begin{abstract}
The physics potential of low-performance and high-performance 
neutrino factories is briefly reviewed.

\end{abstract}



%\maketitle

\section{Introduction}

The recent evidence for neutrino oscillations~\cite{nuosc} 
opens a new and exciting 
era in neutrino physics. We now know that neutrinos of one 
flavor can transform themselves into neutrinos of a different flavor. 
The atmospheric-- and 
solar--neutrino results from the Super-Kamiokande (Super-K) and 
Sudbury Neutrino Observatory (SNO) experiments suggest  
that all three known 
flavors participate in neutrino oscillations. Within the framework of 
three-flavor mixing, the oscillation probabilities are determined by three 
mixing angles ($\theta_{12}, \theta_{23}$, $\theta_{13}$), one 
complex phase ($\delta$), and two mass splittings 
($\Delta$m$^2_{32}$ and $\Delta$m$^2_{21}$ where 
$\Delta$m$^2_{ij} \equiv$ m$^2_i$ - m$^2_j$, the difference between the 
squares of the masses of the neutrino mass eigenstates). The Super-K and 
SNO data suggest that 
(i) $|\Delta$m$^2_{32}| \sim 2 \times 10^{-3}$~eV$^2$,
(ii) sin$^2 2\theta_{23} \sim 1$, and if the LMA MSW solution is confirmed,
(iii) $\Delta$m$^2_{21} \sim 5 \times 10^{-5}$~eV$^2$, and 
(iv) sin$^2 2\theta_{12} \sim 0.87$. 
In addition, the CHOOZ reactor $\nu_e$ disappearance search result implies 
(v) sin$^2 2\theta_{13} < O(0.1)$. However, there is a lot we don't know:  
\begin{itemize}
\item Does three-flavor mixing provide the right framework, or are 
there also contributions from additional sterile neutrinos, neutrino decay, 
CPT-violation ... ?
\item Is sin$^2 2\theta_{13}$ small, tiny, or zero ?
\item Is $\delta$ non-zero ? Is there CP-violation in the lepton sector, 
and does it contribute significantly to baryogenesis via leptogenesis ?
\item What is the sign of $\Delta$m$^2_{32}$ (which determines the neutrino 
mass hierarchy) ?
\item Is sin$^2 2\theta_{23}$ maximal (= 1) ?
\end{itemize}
The answers to these questions may lead us towards an understanding 
of the origin of flavor. However, getting the answers will require 
the right tools, and a neutrino factory~\cite{geer} 
appears to be the tool that we will ultimately require.

\section{Beam properties}

In a neutrino factory muons are stored in a ring with long straight sections. 
Muon decays generate a neutrino beam downstream of each straight section. 
If $\mu^+$ are stored $\mu^+ \rightarrow e^+\nu_e\overline{\nu}_\mu$ decays 
generate a beam consisting of 50\% $\nu_e$ and 50\% $\overline{\nu}_\mu$. 
If $\mu^-$ are stored the beam consists of 
50\% $\nu_\mu$ and 50\% $\overline{\nu}_\e$. 
Design studies~\cite{design} 
suggest that neutrino factories can provide O(10$^{20}$) 
useful muon decays per year. Since the kinematics of muon decay is well known, 
we expect minimal systematic uncertainties on the neutrino flux and 
spectrum. Hence, compared to conventional neutrino beams made using a  
$\pi^\pm$ decay channel, neutrino factories provide $\nu_e$ and 
$\overline{\nu}_e$ beams in addition to $\nu_\mu$ and 
$\overline{\nu}_\mu$ beams, with small systematic uncertainties on the 
beam flux and spectrum. 
%In addition,  
%if the stored muons have energies exceeding about 20~GeV, neutrino 
%factories provide would much higher 
%beam intensities than imagined for future convention neutrino ``superbeams''.

\subsection{$\nu_\mu$ beam properties}
At a high performance 20~GeV neutrino factory providing $2 \times 10^{20}$ 
useful muon decays/yr the $\nu_\mu$ beam flux is about an order of magnitude 
larger than the anticipated future NuMI beam flux at 
Fermilab. Eventually superbeams (very high intensity conventional 
neutrino beams driven by MW-scale primary proton beams) 
may also be able to achieve beam fluxes that are about a factor of 10 
greater than the NuMI flux. However, 
with higher energy neutrino factories the event rate at a distant detector 
increases like $E^3$, rapidly exceeding any corresponding rate we can imagine 
at a superbeam. This is not the whole story. The beam energy 
distributions are also different. 
The neutrino factory beam has a  sharp cut-off 
at the energy of the stored muons. In a conventional neutrino beam there is 
an annoying high-energy tail which gives rise to backgrounds from 
neutral current (NC) events in which a leading 
$\pi^0$ is misinterpreted as an 
electron, faking a $\nu_\mu \rightarrow \nu_e$ signal. 
This background source is absent at a neutrino 
factory. 

\subsection{$\nu_e$ beam properties}
Although the $\nu_\mu$ beam properties are 
interesting, the main reason for wanting a neutrino factory is 
that it would also provide $\nu_e$ and $\overline{\nu}_e$ beams, 
enabling very sensitive searches for $\nu_e \rightarrow \nu_\mu$ 
oscillations. The resulting $\nu_\mu$ component can interact in the 
far detector via the charged current (CC) interaction to produce a muon with 
a charge of opposite sign to that of the muons stored in the neutrino 
factory. The experimental signature is therefore the appearance of a 
wrong-sign muon, for which backgrounds are expected to be at the $10^{-4}$ 
level or lower. In contrast to this, the equivalent 
$\nu_\mu \rightarrow \nu_e$ oscillation search 
using a superbeam suffers from 
background levels that are at about the $10^{-2}$ level. To compare signal 
and background rates it is useful to 
consider some explicit examples~\cite{lindner} 
corresponding to entry-level and 
high-performance neutrino facilities:
\begin{itemize}
\item{Entry-Level Superbeam (JHF $\rightarrow$ Super-K).} 
Running period = 5 years, detector mass = 22.5~kt, proton beam power = 0.75~MW.
\item{High-Performance Superbeam (Super-JHF $\rightarrow$ Hyper-K).}  
Running period = 8 years, detector mass = 1000~kt, proton beam power = 4~MW.
\item{Entry-Level Neutrino Factory (NUFACT I): $1 \times 10^{19}$ 
useful muon decays / year at 50 GeV.}
Running period = 5 years, detector mass = 100~kt.
\item{High-Performance Neutrino Factory (NUFACT II): $2.6 \times 10^{20}$ 
useful muon decays / year at 50 GeV.} 
Running period = 8 years, detector mass = 100~kt.
\end{itemize}
For these four examples, the signal and background rates are compared in 
Table~1. 
%where the oscillation parameters have been taken to be the ones 
%listed in the table caption. 
Based on the signal to background (S/B) ratios 
we might expect that the sensitivity at a neutrino factory will 
ultimately be about two orders of magnitude better than at a 
high-performance superbeam. More detailed studies~\cite{lindner} 
suggest that this naive 
conclusion is not far from the truth although the full story is much more 
complicated. 
\begin{table}
\caption{\label{Table 1}Superbeam and neutrino factory event 
rates from ref.~[4]. The event numbers correspond to the scenarios 
described in the text, with  
$|\Delta$m$^2_{32}| = 3 \times 10^{-3}$~eV$^2$, 
$\Delta$m$^2_{21} = 3.7 \times 10^{-5}$~eV$^2$ 
sin$^2 2\theta_{23} = 1$,   
sin$^2 2\theta_{12} = 0.8 $, 
sin$^2 2\theta_{13} = 0.1$, and $\delta$ = 0.
}
\lineup
\begin{tabular}{@{}lcccc}
\br
 &JHF-SK&JHF-HK&NUFACT I&NUFACT II\\
\mr
Signal    &140&13000&1500&65000\\
Background& 23& 2200& 4.2&  180\\
S/B       &  6&   6 &360 &360\\
\br
\end{tabular}
\end{table}

\section{Physics reach}
Neutrino factory data can be 
separated into 6 subsamples with events tagged by the appearance of 
(i) a right-sign muon, 
(ii) a wrong-sign muon, 
(iii) an $e^+$ or $e^-$, 
(iv) a $\tau^+$, 
(v) a $\tau^-$, or 
(vi) the absence of a lepton. 
Measurements can be made with $\mu^+$ and then with $\mu^-$ stored in 
the ring. Hence there are 12 event energy distributions that can be 
simultaneously fit to obtain the oscillation parameters. 
Since neutrino factories provide intense high energy beams, oscillation 
baselines can be long, or very long (thousands of km). With multiple 
experiments, measurements can be checked with a wide range of baselines. 
Recent studies have suggested that this 
wealth of information will be necessary to pin down the 
oscillation parameters and provide sufficient redundancy to ensure we have 
the right oscillation framework.
\begin{figure}
\begin{center}
\epsfxsize190pt
%\vspace{-1cm}
\epsfbox{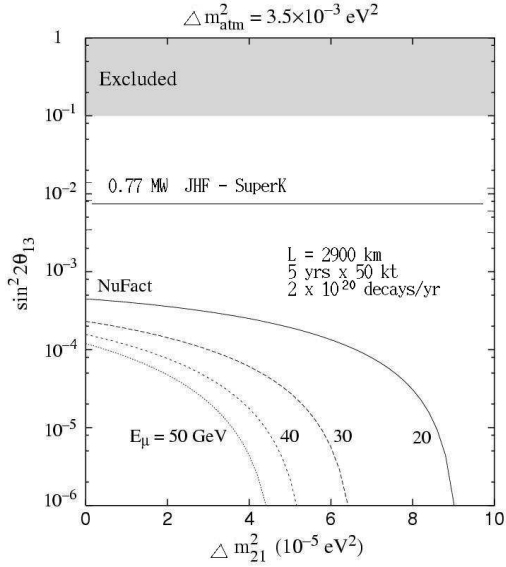}
\end{center}
\caption{\label{label1}
Minimum value of $\sin^2 2\theta_{13}$ for which a 
$\nu_e \leftrightarrow \nu_\mu$ would be observed, shown as a function of 
$\Delta m^2_{21}$ for superbeam and neutrino factory experiments. Figure 
based on calculations described in ref.~6.}
\end{figure}
To better understand this 
it is instructive to expand the expressions for the oscillation 
probabilities in terms of small quantities, keeping just the 
leading order terms. We already know that 
$\sin^2 2\theta_{13}$ is small. We can construct a second small 
quantity $\alpha \equiv \Delta m^2_{21} / \Delta m^2_{31}$. 
Defining $\Delta \equiv \Delta m^2_{31} L / (4E_\nu)$, 
to leading order in $\sin^2 2\theta_{13}$ and $\alpha$, the 
oscillation probabilities in vacuum are given by:
\begin{eqnarray*}
P_{\mu\mu} & = & 1 - \cos^2 \theta_{13} \sin^2 2\theta_{23} \sin^2 \Delta 
 + 2\alpha \cos^2 \theta_{13} \cos^2 \theta_{12} \sin^2 2\theta_{23} 
\Delta \cos \Delta\\
P_{e\mu} & = & \sin^2 2\theta_{13} \sin^2 \theta_{23} \sin^2 \Delta 
 + \alpha^2 \cos^2 \theta_{23} \sin^2 2\theta_{12} \sin^2 \Delta
\\
&& \mp \alpha \sin^2 2\theta_{13} \sin \delta \cos \theta_{13} 
\sin 2\theta_{12} \sin 2\theta_{23} \sin^3 \Delta \\
&& - \alpha \sin^2 2\theta_{13} \cos \delta \cos \theta_{13} 
\sin 2\theta_{12} \sin 2\theta_{23} \cos \Delta \sin^2 \Delta
\end{eqnarray*}
where the $\mp$ sign in the expression for $P_{e\mu}$ corresponds to 
neutrino/antineutrino oscillations. 
Examining the leading order oscillation 
expressions we note that:
\begin{description}
\item{i)} We can replace $\Delta m^2_{31}$ with 
$-\Delta m^2_{31}$ without changing $P_{e\mu}$ 
since vacuum oscillations do not depend upon the 
sign of $\Delta$. For sufficiently long (neutrino factory) baselines within 
the Earth this degeneracy is broken by matter effects.
\item{ii)} For non-maximal mixing we can replace 
$\theta_{23}$ with $\pi/2 - \theta_{23}$ and compensate the change in 
the predicted oscillation probabilities by changing $\sin^2 2 \theta_{13}$. 
\item{iii)} We expect a strong correlation between the fitted values for 
$\delta$ and $\sin^2 2\theta_{13}$. In many cases the best fit combination 
$(\delta, \theta_{13})$ is accompanied by another pair 
$(\delta^\prime, \theta^\prime_{13})$ that yields the same predicted 
leading order oscillation probabilities.
\end{description} 
Hence, we can expect strong correlations between the values of the 
oscillation parameters extracted from fits to the data. 
In addition we can expect degenerate solutions (alternative regions in 
parameter space that are consistent with the data).  
To understand the physics capabilities of neutrino factories (or superbeams) 
we must take into account the impact of these correlations and degeneracies. 
\begin{figure}
\begin{center}
\epsfysize165pt
%\vspace{-1cm}
\epsfbox{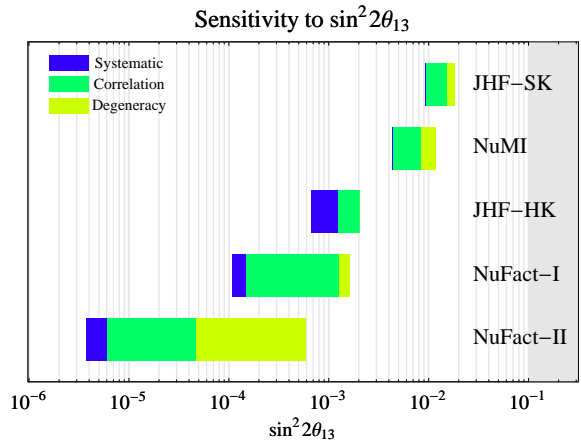}
\end{center}
\caption{\label{label}
Impact of systematics, correlations, and degeneracies on the minimum 
value of $\sin^2 2\theta_{13}$ probed by superbeam experiments (as indicated) 
and neutrino factory experiments with L = 3000~km. Figure from ref.~4.}
\end{figure}
\begin{figure}
\begin{center}
\epsfysize160pt
%\vspace{-1cm}
\epsfbox{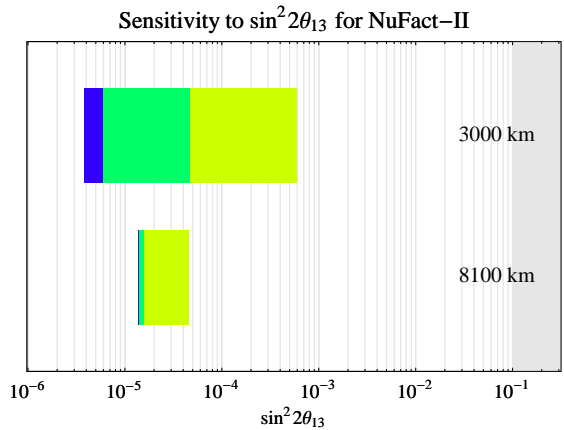}
\end{center}
\caption{\label{label3}
Impact of systematics, correlations, and degeneracies on the minimum 
value of $\sin^2 2\theta_{13}$ probed by neutrino factory 
experiments with baselines of 3000~km and 8100~km. Figure from ref.~4}
\end{figure}

\subsection{$\sin^2 2\theta_{13}$}
Consider first the smallest value of $\sin^2 2\theta_{13}$ that will 
yield a $\nu_\mu \leftrightarrow \nu_e$ signal. 
We will begin by ignoring effects of the all important correlations 
and degeneracies.  
The value of $\sin^2 2\theta_{13}$ that will yield a significant 
appearance signal is shown as a function of $\Delta m^2_{21}$ in Fig.~1 
for the JHF $\rightarrow$ Super-K superbeam and for neutrino factories 
with energies between 20~GeV and 50~GeV. Superbeams can probe values 
of $\sin^2 2\theta_{13}$ an order of magnitude below the present limit. 
If $\Delta m^2_{21}$ is very small, so that the sub-leading scale does 
not contribute to the appearance signal, a high-performance neutrino 
factory would improve on the superbeam sensitivity by more than 
another order of magnitude. If  $\Delta m^2_{21}$ is in the upper half 
of the presently allowed region (spanned by the figure) then neutrino 
factory experiments will measure an appearance signal even if 
$\sin^2 2\theta_{13} = 0$, enabling oscillations generated by the 
sub-leading scale to be directly measured.
\begin{figure}
\begin{center}
\epsfxsize180pt
%\vspace{-1cm}
\epsfbox{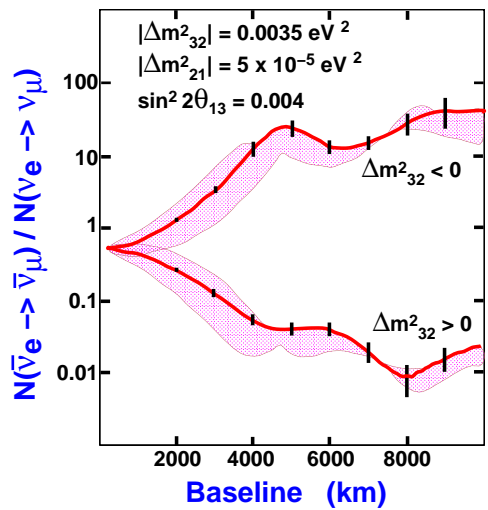}
\end{center}
\caption{\label{label4}
Predicted ratios of wrong-sign muon event rates when $\mu^+$ and $\mu^-$ 
are stored in a 20~GeV neutrino factory, shown versus baseline. The two bands 
correspond to the two signs of $\Delta m^2_{32}$. 
The widths of the bands show the variation as the CP phase 
$\delta$ changes from $-\pi/2$ to $+\pi/2$. The thick lines are for $\delta=0$.
The statistical errors correspond to a neutrino factory providing 
$10^{21}$ muon decays with a 50~kt detector. The figure is from ref.~6.}
\end{figure}

We must now consider the impact of correlations and 
degeneracies. The calculated $\sin^2 2\theta_{13}$ sensitivities (90\% CL) 
from ref.~\cite{lindner} are 
shown in Fig.~2 for the superbeam and neutrino factory scenarios listed in 
Section~2.2. 
The leftmost end of the bars indicate the sensitivities in the 
absence of correlations, degeneracies, and systematic uncertainties. 
The impact of each of these effects on the sensitivity is indicated 
by the shaded sub-bars. Systematic uncertainties degrade the sensitivity 
by a modest amount. With a baseline of 3000~km, 
the impact of correlations and degeneracies limits 
the high-performance neutrino factory 
$\sin^2 2\theta_{13}$ sensitivity to O($10^{-3}$). We can fight 
correlations and degeneracies by using a longer baseline, or  
multiple baselines. In fact with a baseline of $\sim 8000$~km 
the sensitivity is expected to have improved to a few $\times 10^{-5}$ 
(Fig.~3). 
%some two orders of magnitude better than we would 
%expect to achieve at a high performance superbeam. 
%These tentative conclusions 
%need to be confirmed by more detailed studies.
%
\begin{figure}
\begin{center}
\epsfysize150pt
%\vspace{-1cm}
\epsfbox{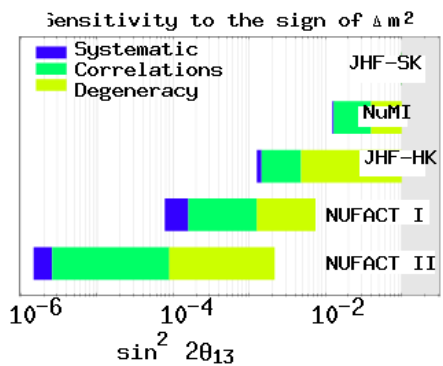}
\end{center}
\caption{\label{label5}
Impact of systematics, correlations, and degeneracies on the minimum 
value of $\sin^2 2\theta_{13}$ for which the sign of $\Delta m^2_{32}$ 
could be determined by superbeam and neutrino factory 
experiments. Figure from ref.~4.}
\end{figure}

\subsection{CP Violation and the neutrino mass hierarchy}
If present, CP violation (CPV) and matter effects will modify the measured 
$\nu_e \leftrightarrow \nu_\mu$ oscillation 
probabilities. These modifications are different for neutrinos and 
antineutrinos. The predicted ratio of events 
$N(\overline{\nu}_e \rightarrow \overline{\nu}_\mu) / 
N(\nu_e \rightarrow \nu_\mu)$ at a neutrino factory experiment with equal 
$\mu^+$ and $\mu^-$ running is shown as a function of baseline in Fig.~4. 
With no CPV and no matter effects (L = 0) the ratio is 0.5, 
reflecting the different neutrino and antineutrino cross-sections. 
As L increases the ratio is enhanced (suppressed) by matter effects 
if the sign of $\Delta m^2_{32}$ is negative (positive). At sufficiently long 
baselines the matter effects are much larger than effects due to possible 
CPV (indicated by the bands in the figure). The sign of 
$\Delta m^2_{32}$ and the CP phase $\delta$ can therefore be determined by 
precise measurements of  
$N(\overline{\nu}_e \rightarrow \overline{\nu}_\mu)$ and  
$N(\nu_e \rightarrow \nu_\mu)$. However, in this simple picture we 
have fixed the values of $|\Delta m^2_{32}|$, $\Delta m^2_{21}$, and 
$\sin^2 2\theta_{13}$. If we now allow all of these parameters to vary in 
our fit to the data we must deal will the resulting correlations and 
ambiguities. The impact of these complications on the sensitivities of 
neutrino factory and superbeam experiments can be seen in Fig.~5, which  
shows the minimum values of $\sin^2 2\theta_{13}$ for which the 
sign of $\Delta m^2_{32}$, and hence the neutrino mass hierarchy, 
can be determined. 
%and (ii) maximal CPV can be distinguished from no CPV. 
With a single experiment at one baseline, 
correlations and degeneracies can degrade the expected sensitivities by 
orders of magnitude. It is believed that at a high-performance 
neutrino factory, with two experiments 
having very different baselines, the sign can probably be determined provided 
$\sin^2 2\theta_{13}$ exceeds O($10^{-4}$) or perhaps a few $\times 10^{-5}$. 
Further study is needed to understand this better, and to  
understand how the picture is improved by combining superbeam and 
neutrino factory measurements. The sensitivity to CPV has also been 
studied and found to be very dependent on $\Delta m^2_{21}$. 
If $\Delta m^2_{21} \sim 4 \times 10^{-5}$~eV$^2$, 
in the center of the presently favored LMA parameter space, 
maximal CPV would be observed at a high performance neutrino 
factory if $\sin^2 2\theta_{13}$ exceeds a few $\times 10^{-4}$.
In the next few years KamLAND is expected to improve our knowledge 
of $\Delta m^2_{21}$, allowing us to sharpen our understanding of 
the CPV capabilities of superbeams and neutrino factories.

\subsection{If LSND is confirmed}
If the LSND oscillation result is confirmed, the simple three-flavor mixing 
framework will need to be modified to include, for example,  
additional light neutrinos that are sterile and/or CPT violation. We 
will already have some knowledge of $\nu_\mu \leftrightarrow \nu_e$ and 
$\nu_\mu \rightarrow \nu_\tau$ oscillations. It seems likely that  
there will be a premium on searching for 
and measuring $\nu_e \rightarrow \nu_\tau$ oscillations, a program 
unique to neutrino factories. It has been shown~\cite{short} 
that there are viable regions 
of four-neutrino mixing parameter space in which both CPV and thousands of 
$\nu_e \rightarrow \nu_\tau$ events could be seen at a  
neutrino factory delivering only O($10^{18}$) decays/yr. Hence, if the 
LSND result is confirmed, a very low intensity neutrino factory might 
provide a well motivated first (fast and cheap ?) step towards the 
high-performance facility we will ultimately want.

\subsection{Non-oscillation physics}
Finally, we must not forget the extensive non-oscillation physics program 
at a neutrino factory facility. 
A high-performance 50~GeV neutrino factory can provide 
$10^6 - 10^7$~ neutrino events per kg per year, enabling highly instrumented 
detectors to obtain data samples of unprecedented magnitude. 
Experiments that might benefit from these intense beams include 
(i) precise neutrino cross-section measurements, 
(ii) structure function measurements 
(with no nuclear corrections), in which 
individual quark-flavor parton distributions can be extracted, 
(iii) precise $\alpha_S$ measurements from non-singlet structure functions, 
(iv) studies of nuclear effects (e.g. shadowing) separately for valence 
and sea quarks, 
(v) spin structure functions, 
(vi) tagged single charm meson and baryon production ( a 1 ton detector 
could yield $10^8$ flavor-tagged charm hadrons / yr), 
(vii) electroweak tests ($\sin^2 \theta_W$ and $\sigma(\nu - e)$),
(viii) exotic interaction searches, 
(ix) neutral heavy lepton searches, and 
(x) searches for anomalous neutrino interactions in EM fields. 
This is only a representative list, to which would be added the 
possible physics programs that exploit very intense cold muon beams 
and the very intense primary proton beams that would also be available 
at a neutrino factory complex.

\section{Summary}

Neutrino factories seem to offer a way to probe $\sin^2 \theta_{13}$ 
and determine the sign of $\Delta m^2_{32}$ provided $\sin^2 \theta_{13}$ 
exceeds a few $\times 10^{-5}$. Neutrino factories would 
also enable $\nu_e \rightarrow \nu_\tau$ oscillation searches. No other 
candidate future facility has these capabilities. Should the 
LMA solar neutrino solution be confirmed, sensitive CPV searches would 
also be 
possible. This would either extend the reach that would already have been 
obtained at superbeams, or possibly follow up an initial indication of CPV 
with a precise measurement of the phase $\delta$. In addition, 
an extensive non-oscillation physics program would enable neutrino factories 
to serve a broad community.

\section*{Acknowledgements}
Our understanding of the physics capabilities of neutrino factories is 
based on the work of many groups over several years. This brief summary 
is indebted to all those that have contributed. I 
am particularly indebted to M.~Lindner for Figs. 2,3, and 5.

\end{document}